# Possible depth-resolved reconstruction of shear moduli in the cornea following collagen crosslinking (CXL) with optical coherence tomography and elastography.

GABRIEL REGNAULT,[1*] MITCHELL A. KIRBY,[1] RUIKANG K. WANG[1,2], TUENG T. SHEN,[2,3] MATTHEW O'DONNELL[1] AND IVAN PELIVANOV[1]

[1]*Department of Bioengineering, University of Washington, Seattle, USA.*
[2]*Department of Ophthalmology, University of Washington, Seattle, USA.*
[3]*School of Medicine, University of Washington, Seattle, USA.*
**gregnaul@uw.edu*

**Abstract:** Corneal collagen crosslinking (CXL) is commonly used to prevent or treat keratoconus. Although changes in corneal stiffness induced by CXL surgery can be monitored with non-contact dynamic optical coherence elastography (OCE) by tracking mechanical wave propagation, depth dependent changes are still unclear if the cornea is not crosslinked through the whole depth. Here, phase-decorrelation measurements on optical coherence tomography (OCT) structural images are combined with acoustic micro-tapping (AμT) OCE to explore possible reconstruction of depth-dependent stiffness within crosslinked corneas in an *ex vivo* human cornea sample. Experimental OCT images are analyzed to define the penetration depth of CXL into the cornea. In a representative ex vivo human cornea sample, crosslinking depth varied from ~ 100 μm in the periphery to ~ 150 μm in the cornea center and exhibited a sharp in-depth transition between crosslinked and untreated areas. This information was used in an analytical two-layer guided wave propagation model to quantify the stiffness of the treated layer. We also discuss how the elastic moduli of partially CXL-treated cornea layers reflect the effective engineering stiffness of the entire cornea to properly quantify corneal deformation.

## 1. Introduction

The cornea is the primary optical element focusing light onto the retina. It contains multiple layers, including epithelium and stroma (Figs. 1a, b). The first acts as a barrier against the external environment and the latter maintains stiffness, transparency and focusing power [1,2]. The microstructure of the stroma is composed of collagen fibrils, arranged in lamellae, lying within a protein rich, hydrated proteoglycan mesh [3,4] (Figs. 1b, c).

Corneal diseases (such as keratoconus (KC)) and surgical complications from refractive surgeries (such as Laser-Assisted In Situ Keratomileusis (LASIK)) may deform the cornea (ectasia) and alter vision. The prevalence of KC in the general population is estimated to be 1.38 per 1000 [5], and nearly 1 million refractive surgeries are performed each year in the USA. Despite their overall success, however, suboptimal visual outcomes and post-refractive corneal decompensation cannot always be predicted for an individual patient.

Corneal collagen crosslinking is a minimally invasive procedure that can potentially slow the progression of corneal ectasia [6–9]. Ultraviolet (UV) light modifies the microstructure of the cornea soaked in riboflavin and forms additional chemical bonds between collagen fibers in the stroma [10] (Fig. 1b). Post-treatment corneas become stiffer and more resistant to enzymatic digestion [11–13]. Although corneal topography (curvature) and thickness maps can be obtained preoperatively, and refractive corrections can be estimated, there is an unmet need to predict corneal decompensation from interventions such as LASIK and CXL therapies. Unfortunately, surgical planning cannot be customized and outcomes (*e.g.,* postoperative corneal ectasia risks) cannot be accurately predicted without quantitatively mapping corneal elasticity. Thus, methods to quantitatively reconstruct corneal elastic moduli are needed.

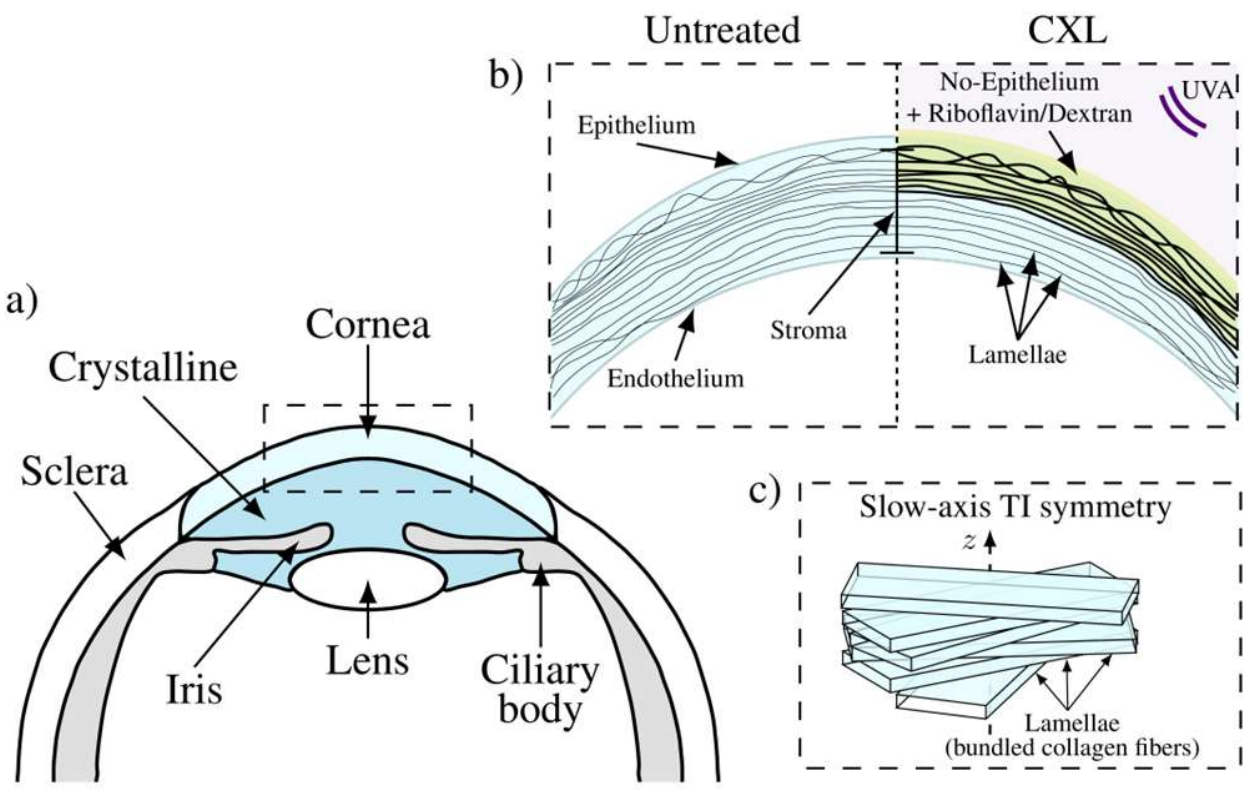

Fig. 1 Illustration of the anterior section of the human eye (a), the change in corneal geometry induced by CXL (b), and three-dimensional organization of corneal lamellae (slow-axis NITI symmetry, i.e., symmetry across the lamellae) within the stroma (c).

Ocular response analyzer (ORA - Reichert Technologies) and Dynamic Scheimpflug Analyzer (DSA - Corvis ST – Oculus Opitkgerate GmbH) are the state-of-the-art in clinical measurements of corneal mechanics. They estimate stiffness as the pressure at inward applanation divided by corneal displacement [14–16]. However, measurements induce large corneal deformations that are often clinically unacceptable, require a non-trivial IOP correction in simulations [17] and assume a simple isotropic mechanical model leading to high variability with experimental conditions. Results obtained with the Corvis ST on KC may be contradictory, and some even show no significant change in corneal stiffness pre- and post-CXL surgery [18,19]. In addition, the result is averaged over the entire cornea with no spatial resolution, and the reconstruction is questionable if corneal thickness varies.

Dynamic elastography is a promising tool to probe soft tissue biomechanics. A shear wave can be launched using direct contact excitation [20–23] or radiation force-based techniques [24,25]. By tracking shear wave propagation, either using magnetic resonance imaging (MRI) [26,27] ultrasound [20–22,28] or dynamic phase-sensitive OCT [29–31], one can infer, with an appropriate mechanical description, the linear [32–35], or non-linear [36,37] stiffness moduli of the tissue. Optical coherence elastography (OCE) is particularly suited to probe corneal biomechanics non-invasively in a clinical environment [24,30,38–40], as it can be combined with non-contact excitation techniques (for example, using an air-puff or non-contact acoustic micro-tapping (AμT [24])).

Because cornea is thin and bounded between air and aqueous humor, wave propagation within it is guided, leading to strong geometric dispersion [33,41]. As such, the common approach associating the Rayleigh surface wave group velocity to stiffness [24,30,38–40,42–45] is not appropriate. In addition, the stroma contains collagen lamellae running in-plane across its width. Lamellae make up approximately 90% of tissue thickness and account for most of the cornea's mechanical structure. They are stacked vertically in approximately 200-500 separate planes [46,47], suggesting an anisotropic mechanical behavior with very different responses to in-plane versus out-of-plane loads (Fig. 1c).

To account for this specific architecture we introduced a model of a nearly-incompressible transverse isotropic (NITI) medium [33], in which corneal stiffness is defined by two (in-, $\mu$, and out-of-plane, $G$) shear moduli, decoupling tensile/inflation properties from shear responses.

Based on this model, we developed an algorithm utilizing guided mechanical waves in a bounded NITI medium to reconstruct both moduli from AμT-OCE. The model was confirmed *ex vivo* in rabbit [48], porcine [49] and human [32,33] models and *in vivo* with rabbit models [48].

For *ex vivo* human corneas, we evidenced that both in- and out-of-plane post-CXL corneal shear moduli experienced an averaged two-fold increase in Young's modulus and an almost four-fold increase for the out-of-plane shear modulus $G$ [32], assuming the whole thickness was treated. That confirmed that CXL increases inter-corneal lamellae crosslinks but less affects corneal deformational properties, defined by the Young's modulus which increases less and has more implications for potential refractive changes.

Despite the success in quantifying CXL-induced corneal elastic properties, there is a common situation where the method [32] must be refined. For example, crosslinking is often inhomogeneous in depth due to more pronounced riboflavin penetration in the corneal anterior where the solution is applied [50], and because UVA irradiation attenuates and is less effective as it propagates through cornea [51]. Together, this generally produces a clear demarcation line between treated and untreated regions [52], suggesting a two-layer structure postoperatively.

As noted above, mechanical waves generated in the cornea are typically guided. They occupy the entire thickness of the cornea and, therefore, carry depth-accumulated information. Thus, reconstructing the depth dependence of corneal moduli is difficult with wave based OCE alone without a good estimate of the CXL penetration depth into the cornea.

Blackburn et al. [53] recently introduced a novel metric to track CXL penetration within the cornea using time-resolved OCT. They demonstrated that the phase decorrelation decay rate of the complex OCT signal is reduced in the CXL area, which can be used to distinguish treated from untreated areas postoperatively.

In this paper, we combine the method described in [53] with AμT-OCE measurements to explore possible reconstruction of both in- and out-of-plane corneal elastic moduli over depth in a partially CXL-treated ex vivo human cornea. We developed an analytical model of guided wave propagation accounting for multiple layers, each with distinct stiffness moduli and thickness, to properly account for CXL-induced corneal layering. Baseline elastic moduli in an untreated *ex vivo* cornea can be determined from AμT-OCE prior to CXL, so that induced changes in the treated anterior layer can be quantified by fitting the post-CXL wave dispersion dependence in the frequency-wavenumber ($f$-$k$) domain. We also discuss how the depth-distribution of stiffness affects the effective engineering stiffness of the entire cornea and show that assessing stiffness in both layers is needed to properly predict corneal deformation and quantify surgical outcomes.

## 2. Method

### *2.1 Cornea preparation*

A corneal-scleral ring stored in Optisol (Chiron Ophtalmics) was obtained from CorneaGen. It came from a 26 year-old donor and was stored for less than 30 days. The corneal-scleral button was mounted on an artificial anterior chamber (Barron, CorzaMedical; see Fig. 2), connected through the inlet port to an elevated bath filled with balanced saline solution (BSS) to apply a controlled pressure mimicking intraocular pressure (IOP) on the anterior segment of the cornea. The outlet port remained closed to allow the IOP to settle at 15 mmHg within the chamber, corresponding to human physiological conditions [54]. CXL followed the Dresden protocol [6]. First, the epithelial membrane was removed. Then, the cornea was soaked in riboflavin for 30 minutes by applying a 50 μL drop of 0.1% riboflavin in 20% dextran solution every two minutes. It was then exposed to 3 mW/cm$^2$ of 370 nm UV light for 30 minutes, while a drop was re-applied every 5 minutes.

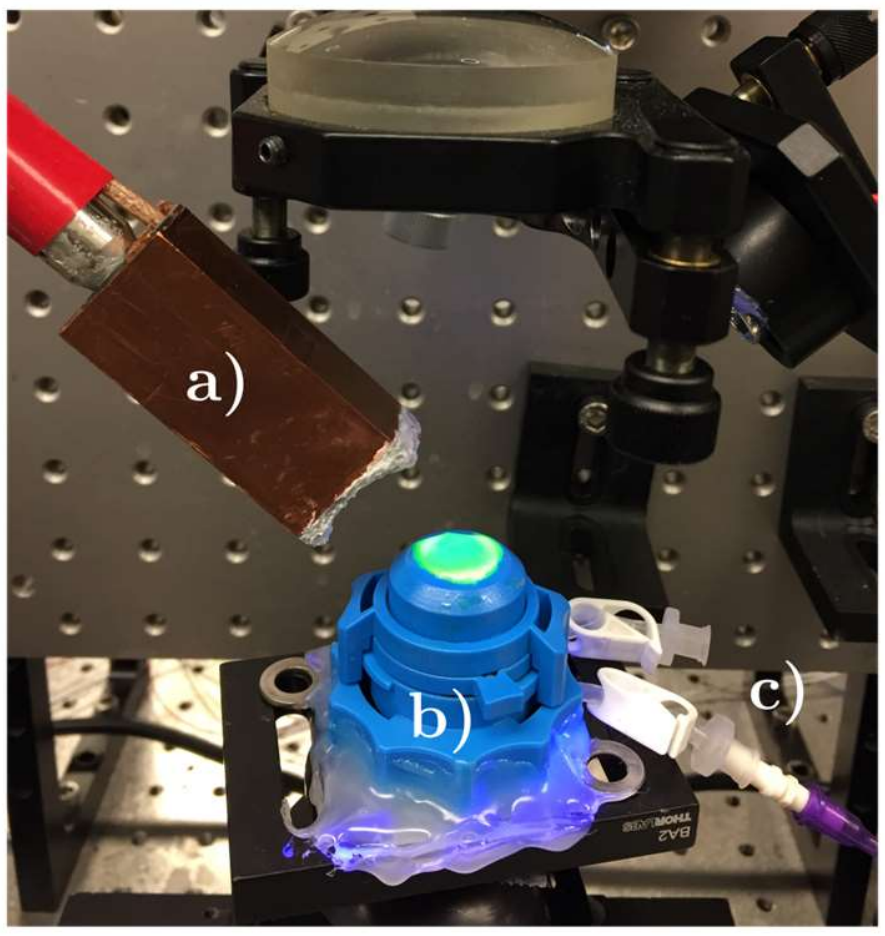

Fig. 2. Picture of the experimental set up during UV-CXL. a) Acoustic micro-tapping transducer. b) Artificial anterior chamber with c) inlet port connected to the elevated bath and outlet port closed to control IOP.

### 2.2 AμT-OCE imaging system

A cylindrically focused air-coupled (AμT) transducer, operating at a 1 MHz frequency, launched mechanical waves in the cornea. A spectral domain OCT system with a 46.5 kHz A-scan rate was used to track wave propagation and structural changes [24,30,32,33]. The cylindrical focus generated quasi-planar guided waves within the cornea. The OCT system operated in M-B mode, where a single push was triggered by the system while 512 consecutive A-scans were taken at a fixed location (M-scan). The M-scan sequence and push excitation were repeated for 256 locations, creating a volume with 256 x-samples, 1024 z-samples and 512 t-samples (see Fig. 3a)), with an effective imaging range of 6 mm × 1.2 mm×11 ms.

The particle velocity along the probe beam direction was obtained from the optical phase difference between two consecutive A-lines at each location [55]. The spatio-temporal ($x$-$t$) surface signature of the guided wave was computed from an exponentially weighted-average of the particle velocity over the first 180 μm. As shown in Fig. 3b, the guided wave only propagated during the first 4 ms of the scans, which was used to determine the stiffness of the material by fitting the computed dispersion curve in the frequency-wavenumber domain ($f$-$k$) obtained from the 2D Fourier spectrum (Fig. 3c). This procedure is detailed in Section 2.4. On the other hand, data from the last 7 ms were used to study structural changes with phase decorrelation [53] (see Section 2.6).

### 2.3 Multi-layer NITI model

Like most biological tissue, the cornea is nearly incompressible. In addition, its structure implies a transverse isotropy [49] and, therefore, its mechanical behavior under small deformation should be described with a NITI model [33]. In Voigt notation, Hook's law of stress and strain for a NITI material takes the form:

$$\begin{bmatrix}\sigma_{xx}\\ \sigma_{yy}\\ \sigma_{zz}\\ \tau_{yz}\\ \tau_{xz}\\ \tau_{xy}\end{bmatrix}=\begin{bmatrix}\lambda+2\mu & \lambda & \lambda & & & \\ \lambda & \lambda+2\mu & \lambda & & & \\ \lambda & \lambda & \lambda+\delta & & & \\ & & & G & & \\ & & & & G & \\ & & & & & \mu\end{bmatrix}\begin{bmatrix}\epsilon_{xx}\\ \epsilon_{yy}\\ \epsilon_{zz}\\ \gamma_{yz}\\ \gamma_{xz}\\ \gamma_{xy}\end{bmatrix}, (1)$$

where $\sigma_{ij}$ denotes engineering stress, $\epsilon_{ij}$ denotes engineering strain, $\tau_{ij}$ denotes shear stress, $\gamma_{ij} = 2\,\epsilon_{ij}$ denotes shear strains, the subscripts $x$, $y$ and $z$ refer to the Cartesian axes and $G$, $\mu$, $\lambda$ and $\delta$ are four independent elastic constants. In previous studies [34,49], we have demonstrated that $\delta$, which accounts for tissue tensile anisotropy, cannot be determined from the propagation of vertically polarized guided waves generated in the cornea. At the same time, the influence of $\delta$ on the in-plane Young's modulus, $E_T$, is minor so that it is restricted to the range of $2\mu \leq E_T \leq 3\mu$. Consequently, here, corneal tensile isotropy ($\delta = 0$) is also assumed ($E_T = E \cong 3\mu$).

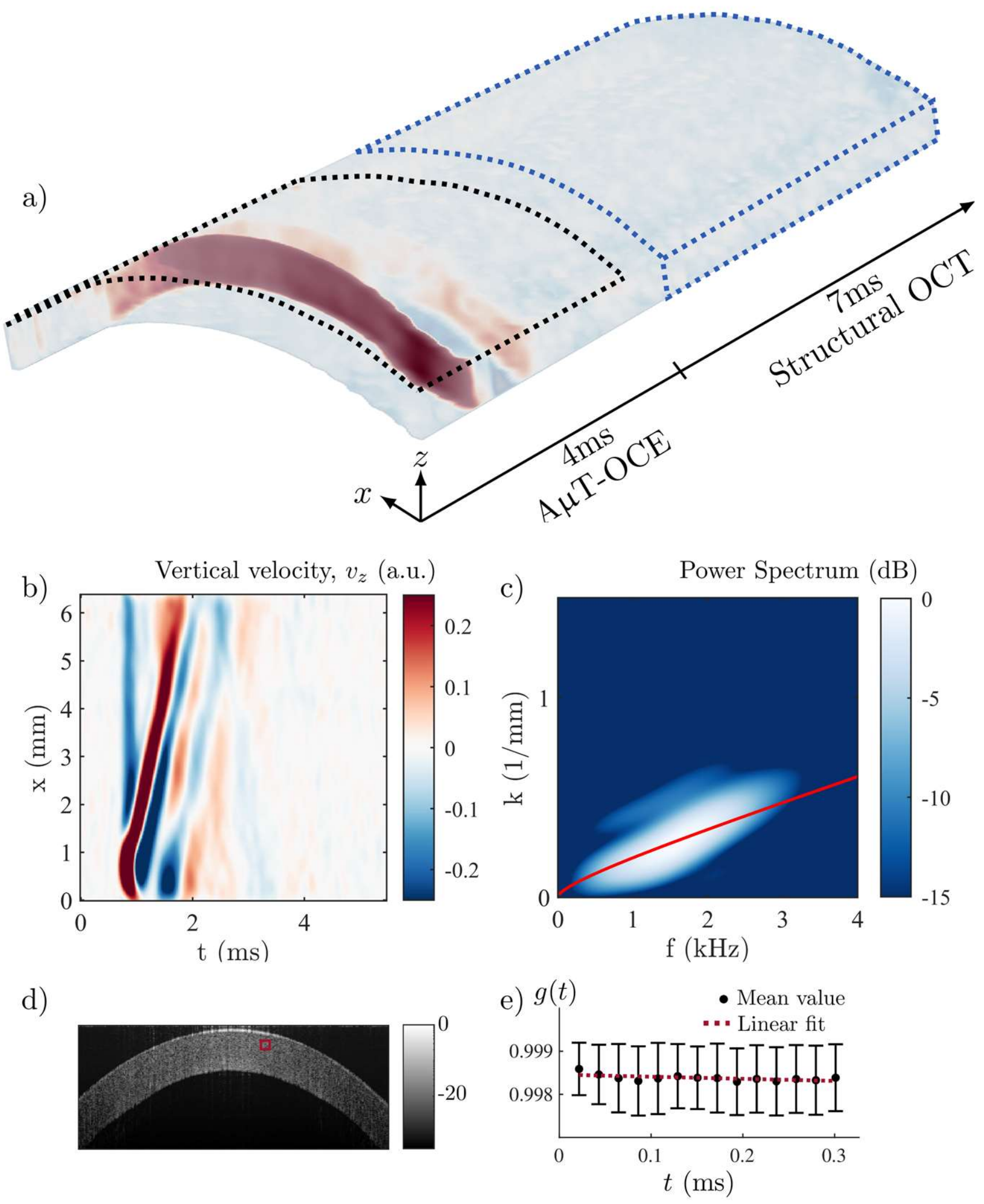

Fig. 3. Diagram of spectral domain time-resolved OCT and AμT-OCE measurements in a pre-CXL cornea. a) 3D ($x, z$ and $t$), wavefield after AμT excitation. The top surface wavefield of the initial time sequence (black dotted region) is used for elastic moduli reconstruction, and data at the end of the sequence are used for phase decorrelation measurements. b) $x$-$t$ plot showing the top surface signature of the guided mode during the first 5 ms of the acquisition sequence. c) $f$-$k$ spectrum obtained by 2D-FFT of the $x$-$t$-plot showing the dispersion dependence of the first anti-symmetric mode $A_0$. The red curve indicates the best fit obtained with the NITI model [33]. d) Structural OCT image obtained by averaging the last 7 ms of the raw OCT signal. e) Phase decorrelation function $g(\tau)$ at the location indicated by the red square in d).

Because the cornea is a nearly incompressible soft tissue, its Young's modulus does not depend on $\lambda$. Therefore, among the four elastic constants, only $G$ and $\mu$ (respectively the out-of-plane and in-plane shear moduli) are needed to predict corneal deformation under mechanical loading.

The effects of CXL on the cornea depend on depth. Several recent studies showed that postoperative CXL corneas might experience non-uniform crosslinkage with depth. The transition between crosslinked (anterior) and non-crosslinked (posterior) parts tends to be sharp rather than smooth [51,52]. This effect is also observed in our experiments (see below). Thus, a two-layer structure is considered an appropriate model to quantify postoperative corneas.

CXL was also shown to change collagen fiber diameter and interfibrillar spacing [50], but nothing suggests a modification of its macroscopic anisotropic organization. Based on this observation, we developed a multi-layer model to predict wave propagation within CXL corneas (see Supplementary 1) that accounts for any arbitrary number of layers, each with a stress-strain relationship given by Eq. (1) and linked by solid-solid boundary conditions (continuity of normal components of stress and displacement across every interface). Accounting for the external boundary conditions (liquid below and air above the cornea) and the finite thickness of the medium, the dispersion relation for guided waves can be determined directly from stiffness moduli $G_n$ and $\mu_n$ and the thickness $h_n$ of each layer. A more detailed description of the multi-layer model is in Supplementary 1. Although only 2 layers were considered here, the multi-layer model can be used to accurately capture more complicated transitions between crosslinked and non-crosslinked areas.

In an untreated cornea, only the first anti-symmetric mode, referred to as $A_0$, typically propagates in the range of frequencies that can be recorded in OCE (usually <5 kHz). Because a partial-CXL cornea consists, in our approximation, of two horizontally assembled layers, each having a vertically aligned symmetry axis, this symmetry holds for the global material. Thus, only the $A_0$-mode is also expected in the partially crosslinked cornea.

### *2.4 Fitting $f$-$k$ spectra pre- and post-CXL*

Prior to treatment, the cornea was assumed homogeneous, which in our model corresponds to a single NITI layer bounded above by air and below by water. The experimental $f$-$k$ spectrum (see Fig. 3c) was obtained by computing the 2D FFT of the $x$-$t$ plot. Shear moduli $G$ and $\mu$ in pre-CXL cornea were obtained by fitting the measured $f$-$k$ spectrum with the analytic dispersion relation for the $A_0$-mode [32,33,48,49].

In the CXL-treated cornea, the thickness of both layers can be measured (see Section 3.1), and the posterior layer is assumed to still possess the original (*i.e.*, untreated) elastic properties. Thus, the 2-layer model with known elastic moduli of the bottom (untreated) layer can be used to determine the stiffness of the top layer.

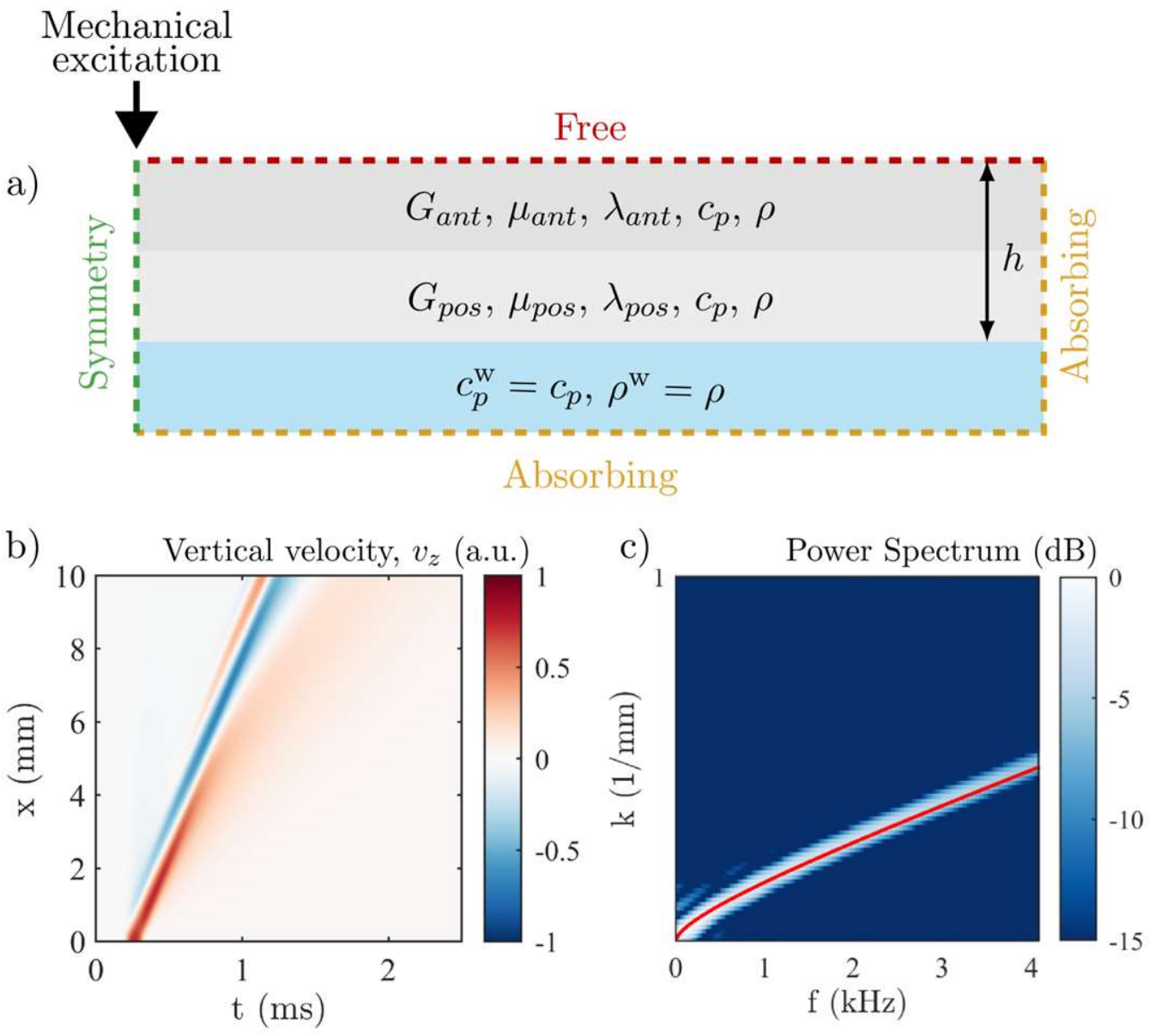

Fig. 4. Finite element simulations to study the effects of a layered structure for post-CXL cornea. a) Geometry of the two-layered material used in simulations, bounded above by air and below by water. b) Top surface spatio-temporal signature ($x$-$t$ plot) of the guided wave for a two-layer case with $G_{ant} = 296.8$ kPa, $\mu_{ant} = 34.6$ MPa, $G_{pos} = 59.5$ kPa, $\mu_{pos} = 7.3$ MPa, $h_{ant} = 150$ µm and $h_{pos} = 370$ µm. c) 2D Fourier spectrum of the wave studied in b) showing the main propagating $A_0$-mode and, in red, the analytical solution obtained from the multi-layer NITI model with identical parameters.

To ensure reliable fitting for all cases, we computed a goodness of fit (GOF) metric $\Phi = \frac{\sum_f \chi_{fit}(f)}{\sum_f \chi_{max}(f)}$, where $\chi_{fit}(f)$ corresponds to the energy of the 2D spectrum covered by the best analytical solution (one or two layers) at a given frequency $f$ and $\chi_{max}(f)$ is the unconstrained maximal energy of the spectrum at frequency $f$. Based on recent results (see Supplemental Material in [32]), reliable fitting in human *ex vivo* corneas is associated with values of $\phi > 0.9$. An example of a 2D-spectrum and the fitted $A_0$-mode obtained with this procedure for the untreated case is shown in Fig. 3c.

### *2.5 FEM simulations*

We designed finite element (FEM) simulations in OnScale to determine the accuracy of our multi-layer NITI model in reconstructing stiffness along corneal depth. The geometry is shown in Fig. 4a. Corneal boundary conditions were replicated, with the material bounded above by air and below by water. The speed of sound in all layers (material and water) was fixed to avoid the reflection of compressional waves at the air-liquid and inter-layer boundaries. It also improved the absorption of compressional waves at the absorbing boundaries and, thus, avoided divergent simulations. The simulated transient excitation of broadband elastic waves closely matched that used in AµT experiments. More details about the simulations can be found in [33].

Based on phase-decorrelation measurements (see Section 2.6), we assumed that post-CXL two layers with distinct thicknesses were formed within the cornea, the top layer being stiffer than the bottom one. Stiffness values assessed from experiments were also used in the simulations. We used the top surface vertical particle velocity of the simulated wave (see Fig. 4b), and its associated $f$-$k$ spectrum (see Fig. 4c), to show that the analytical solution obtained from the N-layer model (see Eq. S39 in Supplementary 1) closely matches the $f$-$k$ spectrum of numerically simulated wave propagation. This confirms the accuracy of the analytical model in quantifying measurements of corneal elastic moduli and their variation with depth.

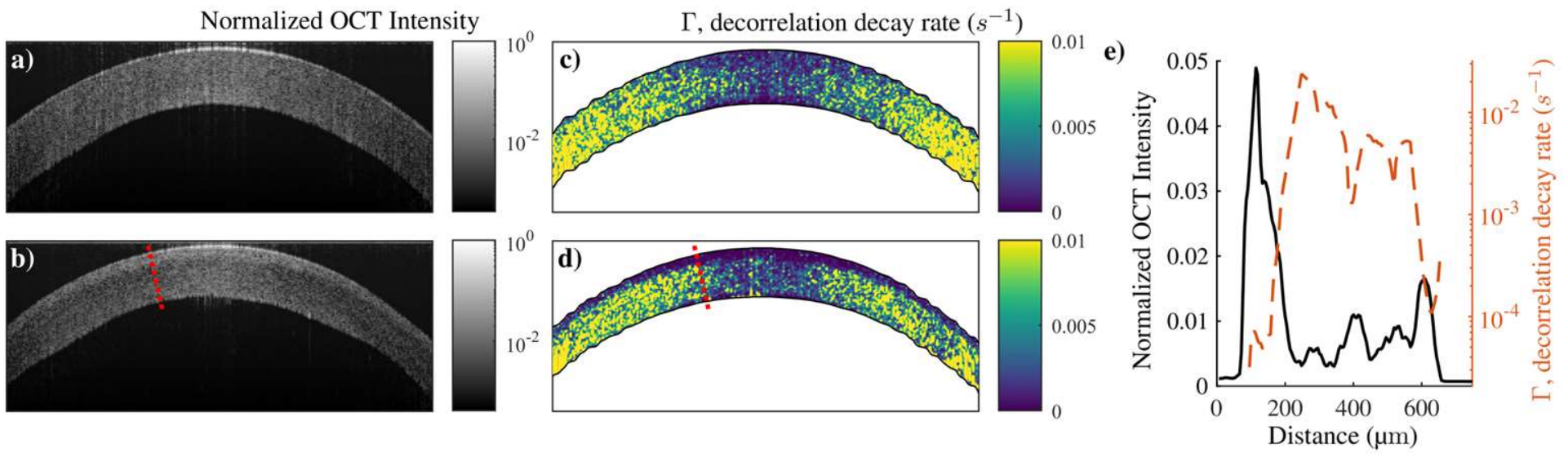

Fig. 5. Short time decorrelation before and after CXL. Structural OCT images averaged over the last 7ms of the OCT scan for a) pre-CXL and b) post-CXL. Maps of decorrelation coefficient Γ for c) pre- and d) post-CXL. e) Profile of OCT intensity and Γ along the red dotted line shown in b) and d) for the CXL cornea.

### 2.6 Phase Decorrelation OCT (PhD-OCT)

Blackburn et al. [53] have recently introduced a metric to track CXL penetration within the cornea using time-resolved OCT. It was shown that the phase decorrelation decay rate of the complex OCT signal is reduced in CXL areas and can distinguish treated from untreated areas after the procedure.

In our study, the autocorrelation function of the signal g(τ) was computed over 15 consecutive samples at 46,500 Hz for six consecutive pixels within a given A-line:

$$g(\tau) = \left\langle \frac{\langle E(t)\, E^{*}(t+\tau)\rangle_{pixels}}{\sqrt{\langle E(t)\, E^{*}(t)\rangle_{pixels}} \times \sqrt{\langle E(t+\tau)\, E^{*}(t+\tau)\rangle_{pixels}}} \right\rangle, \quad (2)$$

which is expected to follow an exponential decay [56]:

$$g(\tau) = e^{-\Gamma . \tau} \approx 1 - \Gamma \cdot \tau\,, \quad (3)$$

where Γ is the decorrelation coefficient that is inversely proportional to the Brownian diffusion coefficient [56], meaning that the more coherent the material, the smaller the decorrelation coefficient. The procedure was performed starting at $n$, $n+1$, $n+2$, … A-lines, where $n$ is the first time-sample used for phase-decorrelation ($t(n) = 4$ ms). The decorrelation coefficient Γ was then computed using the averaged $g(\tau)$ over the number of starting points by fitting with a first order polynomial (see Fig. 3e): $\langle g(\tau)\rangle = b - \Gamma \cdot \tau$, where $\langle\ \rangle$ denotes the average over the number of starting points. In crosslinked regions of the cornea (anterior), tissue stiffens, implying that Γ should be smaller than in the untreated region (posterior). For post-processing, we rejected all fits with $b < 0.95$, corresponding in general to peripheral regions where the signal to noise ratio (SNR) was too low.

## 3. Results

### 3.1 Thickness of CXL layer

The spatial distribution of the OCT intensity signal (Figs. 5a, b, e) and phase decorrelation images (Figs. 5c, d, e) both show clear layering in the treated cornea. The effect of CXL is not homogeneous across the cornea, with a more pronounced effect at the center (~150 μm) than at the periphery (~ 100 μm). For the present case, we estimated that about 30% of the cornea was treated effectively. The treated cornea is thinner than that prior to CXL (its thickness reduced from 575 μm to 520 μm), as generally observed in the literature [57,58].

Table 1. Corneal thickness (h), in- (μ) and out-of-plane (G) elastic moduli and goodness of fit (Φ) pre- and post-CXL

| | $h$ (μm) | $G$ (kPa) | $\mu$ (MPa) | $\phi$ |
|---|---|---|---|---|
| | | **Pre-CXL** | | |
| Fitted moduli | 575 | $59.5 \mp (5{,}8)$ | $7.6 \mp (7{,}13)$ | 0.961 |
| | | **Post-CXL** | | |
| Top layer | 150 | $296.8 \mp (60, 92)$ | $34.6 \mp (20, 23)$ | 0.95 |
| Bottom layer | 370 | $59.5 \mp (5{,}8)$ | $7.6 \mp (7{,}13)$ | - |
| Effective engineering moduli, Eqs. (4), (5) | 520 | $77.3 \mp (6{,}10)$ | $15.4 \mp (8{,}11)$ | - |

### 3.2 Stiffness of CXL-treated corneal layers

AμT-OCE scans, taking approximatively 3 s to acquire and save data, were acquired prior and post-CXL. A space-time (*x-t*) plot of the vertically polarized particle velocity (see Figs. 3a, b) in the untreated cornea was used to compute the $f$-$k$ spectrum (see Fig. 3c), which was then fitted using the procedure detailed in Section 2.4, assuming the cornea as a single homogeneous layer. The fitting routine was also detailed in our recent work [32,48]. Results for the reconstructed in- ($\mu$) and out-of-plane ($G$) corneal shear moduli pre-CXL are $\mu = 7.6 \mp (7{,}13)$ MPa and $G = 59.5 \mp (5{,}8)$ kPa (see Table 1).

To reconstruct depth-dependent stiffness moduli after CXL, it is assumed that: i) the thickness of both the anterior and posterior layers can be measured using dynamic OCT from phase-decorrelation and/or intensity variation methods (see Fig. 5, estimated to be $h_{ant} = 150$ μm and $h_{pos} = 370$ μm with both methods); ii) the stiffness of the posterior layer remains unchanged after CXL; iii) the effect of CXL is homogeneous in the anterior layer. Fixing known parameters (untreated cornea thicknesses and posterior stiffness moduli), stiffness moduli of the anterior cornea layer can be determined by fitting the wave dispersion curve in the $f$-$k$ domain with the 2-layer model (Fig. 6b). We found an increase in both stiffness moduli $G_{ant} = 296.8 \mp (60, 92)$ kPa and $\mu_{ant} = 34.6 \mp (20{,}23)$ MPa compared to those for the posterior region $G_{post} = 59.5 \mp (5, 8)$ kPa and $\mu_{post} = 7.6 \mp (7, 13)$ MPa (see Figs. 6c, d). The goodness of fit for the 2-layer model ($\Phi = 0.950$) remained within the range of reliable fitting. The results are summarized in Table 1.

GOF variation for a given sample at a fixed IOP was previously shown to be about 1% [32]. We used this fact to build error bars for the present study, illustrated in Fig. 6d for the two-layer fitting procedure. First, we fitted the projection of the goodness of fit surface for a value of $\Phi = 0.99 \times max\ (\Phi)$ (*i.e.*, 1% below the optimal GOF), with an ellipsoidal function that best described the shape of the iso-goodness levels. Then, we computed the uncertainty intervals as the intersection of the horizontal and vertical direction with the fitted ellipse. This produces asymmetric uncertainty intervals (particularly for $\mu$), reflecting the asymmetric variation of $\Phi$.

### 3.3 Mixing rules for the effective engineering moduli of layered materials

A theory for effective moduli of multi-layer materials was developed in the early 1970's for composite materials. It is now accepted in material science and broadly used in the development of composite structures [59–61]. The derivation of these 'effective' material engineering moduli is based on 'mixing rules' of stiffness moduli across depth using the following assumptions: i) out-of-plane stresses and in-plane strains are uniform across thickness; ii) in-plane stresses and out-of-plane strains are averaged across thickness based on layer volume fractions. Note that the solution is valid only for low-frequency material deformation, which are, fortunately, usually related to physiologically induced stresses. As such, even if these definitions do not hold for perturbations of any kind, they are appropriate for *in vivo* corneal response to physiological loads.

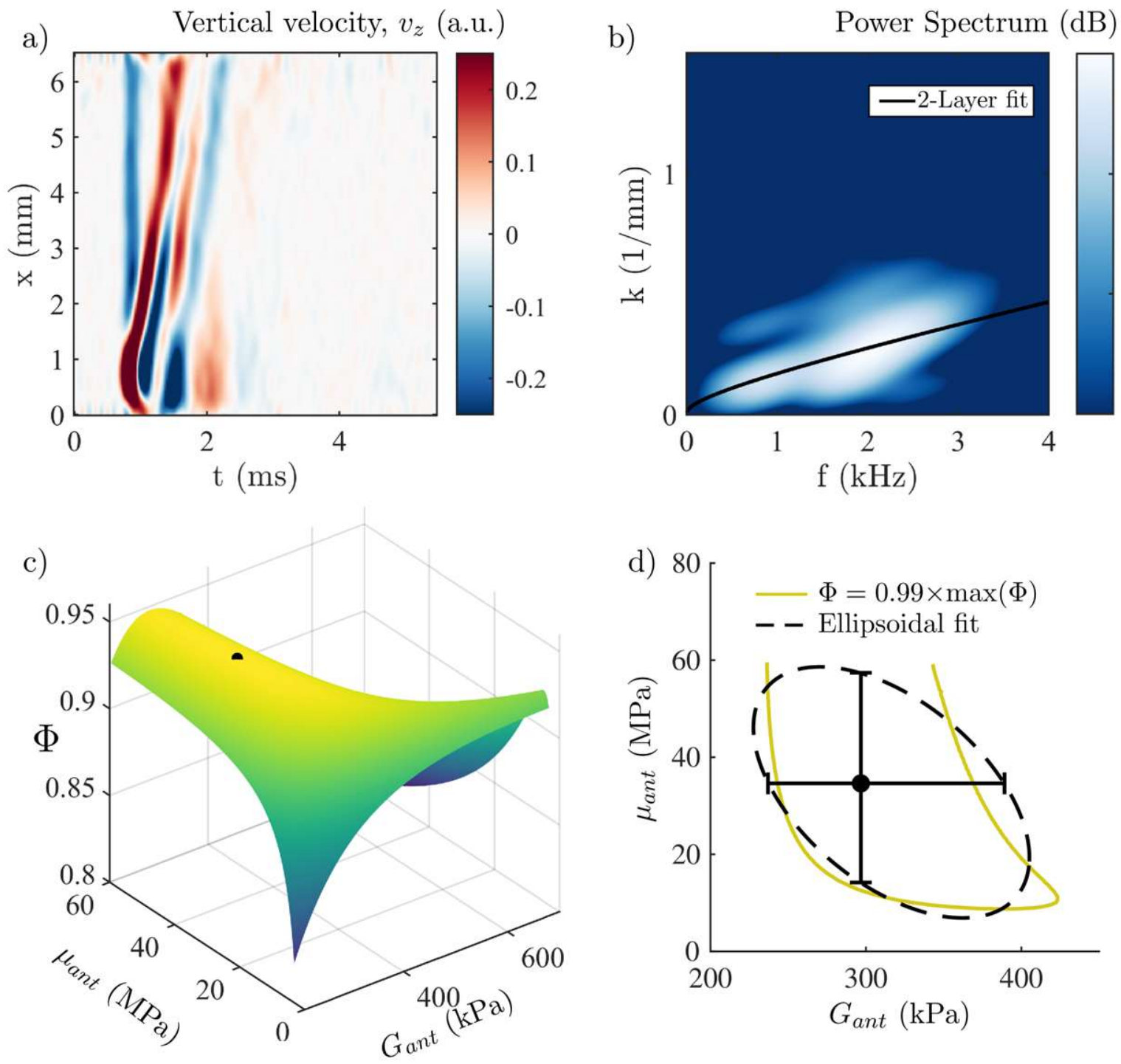

Fig. 6. Post-CXL fitting. a) Measured vertically polarized top-surface signature of the guided wave in the treated cornea. b) 2D-spectrum computed from a). c) 2D Goodness of Fit surface when the fit is performed with the 2-layer model to determine the anterior layer stiffness moduli. d) Projection of the surface plot for $\Phi = 0.99 \times max\ (\Phi)$. Error bars are computed as the intersection of the vertical and horizontal directions with the fitted ellipse. The global maximum of $\Phi$ is indicated in c) and d) by the circular marker.

Sun et al. [60] have demonstrated that an effective out-of-plane engineering modulus $G_{eff}$ can be computed using the inverse mixture rule for out-of-plane material constants of individual material layers:

$$G_{eff} = \left(\sum_n \frac{h_n/h}{G_n}\right)^{-1}, \qquad (4)$$

where $h$ is a total material thickness, $h_n$ is a thickness of the $n^{\text{th}}$ layer and $G_n$ is an out-of-plane modulus of this layer. On the other hand, an effective in-plane engineering modulus $\mu_{eff}$ can be obtained from the mixture rule:

$$\mu_{eff} = \sum_n \mu_n . \frac{h_n}{h}, \qquad (5)$$

where $\mu_n$ is an in-plane modulus of the $n^{\text{th}}$ layer.

Based on the mixture rules described above, the effective low-frequency engineering moduli of a partially treated cornea for our case are computed to be $G_{eff} = 77.3 \mp (6{,}10)$ kPa and $\mu_{eff} = 15.4 \mp$ (8,11) MPa (see Table 1).

## 4. Discussion and conclusions

In this study, we combined structural OCT with dynamic AμT-OCE to assess the penetration depth of CXL in the cornea. Analyzing the brightness of structural OCT images and the rate of image decorrelation between consecutive B-scans, we confirm a sharp transition between CXL and untreated cornea layers and measure the thicknesses of treated and untreated layers. This finding suggests a 2-layer medium model for the treated cornea that can be used to reconstruct both in- and out-of-plane elastic moduli in the treated layer.

As we discussed in detail in our previous work [32,48], the $A_0$-mode dispersion spectrum is much more sensitive to variations of out-of-plane modulus $G$ rather than to variations of in-plane modulus $\mu$. This results in large error bars in the reconstruction of $\mu$ (see Table 1) compared to those for $G$. However, here we used only a single measurement, and reconstruction accuracy can be improved by repeating AμT measurements. Note that the error bar asymmetry comes from the asymmetry of the GOF function (see Figs. 6 c, d)), which was previously described in Refs [32,48].

Using AμT-OCE, we tracked guided wave propagation in the cornea before and after CXL. Using the NITI model in each layer, we quantified depth-localized corneal stiffening with CXL. Because the $A_0$-mode occupies the whole thickness, it carries averaged information about material stiffness (*i.e.*, averaged over its two parts). We determined the effective moduli in the treated corneal layer by fitting the 2D-spectrum with the two-layer model using known thicknesses for each layer and known elastic moduli for the untreated part (obtained from OCE measurements pre-CXL). Effective engineering moduli of the entire cornea can be then calculated using Eqs. (4) and (5), respectively, for out-of- and in-plane moduli.

As explained in Section 3.3, elastic moduli determined for the anterior (CXL-treated) corneal layer can be used to compute effective corneal engineering moduli, where $\mu_{eff}$ uses a simple mixture rule whereas $G_{eff}$ requires an inverse mixture rule. This implies that even if the treated layer $G$ experience an almost six-fold increase, its effective increase for the whole tissue is only by a factor of 1.3. On the other hand, $\mu$ experiences a local five-fold increase in the anterior layer but the overall in-plane modulus, which is more directly related to deformations in response to physiological loads, increases by a factor of 2.

Because the mixture rules for engineering effective moduli (Eqs. (4), (5)) assume only low-frequency perturbations, it is interesting to check if they could also describe guided wave behavior in the partially crosslinked cornea when considered as an effective homogeneous single-layer material (see Supplementary 2). We found that the effective engineering mixture rules cannot be applied to quantify 'effective' guided wave behavior in the layered medium and, more importantly, the guided wave behavior considered in the 'effective' single layer incorrectly describes effective corneal engineering moduli. Fitting the experimentally obtained *f-k* spectrum (Fig. 6b) with the single layer model (Supplementary Fig. S1) results in a different (incorrect) set of reconstructed engineering corneal moduli (see Table 2 below). Thus, measuring the depth of CXL penetration into cornea and implementing the multi-layer guided wave model are both required to accurately assess post-CXL corneal mechanical moduli.

Although the relationship between acoustic bulk (longitudinal and shear) wave propagation speeds and mechanical moduli in multi-layer or multi-component media has been reported in several studies [62,63], these relationships have not been explored for guided waves due to their high geometric dispersion. The situation is even more complicated for anisotropic media. The lack of the complete solution to this problem does not affect the goal of this study and is definitely outside its scope. However, we would like to share an interesting observation. We empirically found that both the analytic model and *ex vivo* experiments in the human cornea sample suggest that a simple mixture rule (Eq. (5)) for both corneal moduli can be applied to approximately compute effective guided wave propagation in a multi-layer NITI medium, as detailed in Supplementary 2.

Table 2. Effective moduli measured with the single or multilayer models.

| | $G$ (kPa) | $\mu$ (MPa) |
|---|---|---|
| 2-layer model, Eqs. (4), (5) | $77.3 \mp (6,10)$ | $15.4 \mp (8,11)$ |
| Single layer model | $127.5 \mp (12,17)$ | $9.3 \mp (8,18)$ |

One of the limitations of this work is that it assumes that post-CXL cornea contains two homogeneous layers, a reasonable assumption given several previous studies [50–52]. However, when a more gradual transition between CXL-corneal layers is observed, the multi-layer model introduced here can be used to compute $A_0$-mode dispersion for more sophisticated models of CXL (*e.g.*, accounting for a gradient in stiffness or more complex structural changes).

Recent results suggest that reverberant OCE can reconstruct depth-dependent stiffness variations [23]. It would be interesting to compare it with our method in future studies. Note, however, that reverberant OCE is not currently feasible *in vivo* because it uses contact vibrators. This is why guided wave-based OCE is still the only method capable of *in vivo* non-contact measurements of corneal anisotropic elasticity, ultimately with sub-mm lateral resolution [64].

Finally, we have shown that phase-sensitive OCT combined with AμT wave excitation can assess both the structure of human cornea and the depth-dependence of moduli due to CXL. These findings are essential for building personalized models of corneal deformation following CXL and, thus, better adapt crosslinking therapy for clinical use and predict its outcomes. Further experiments on a larger group of cornea samples are required to generalize the present results.

**Funding.** NIH Grants R01-EY026532, R01-EY028753, and R01-AR077560 and the Department of Bioengineering at the University of Washington

**Acknowledgments.** TODO

**Disclosures.** The authors declare no conflicts of interest.

**Data availability.**

**Supplemental document.** See Supplement 1 and 2 for supporting content and Supplement 3 for Matlab scripts of the N-layer model.

# Supplementary Material:

## Possible depth-resolved reconstruction of shear moduli in the cornea following collagen crosslinking (CXL) with optical coherence tomography and elastography.

GABRIEL REGNAULT,[1*] MITCHELL A. KIRBY,[1] RUIKANG K. WANG[1,2], TUENG T. SHEN,[2,3] MATTHEW O'DONNELL[1] AND IVAN PELIVANOV[1]

[1]*Department of Bioengineering, University of Washington, Seattle, USA.*
[2]*Department of Ophthalmology, University of Washington, Seattle, USA.*
[3]*School of Medicine, University of Washington, Seattle, USA.*
**gregnaul@uw.edu*

### Supplementary Material 1:

### Guided mechanical wave propagation in a multilayered NITI material

Assume that a CXL-treated cornea can be modelled as a laminate of NITI layers. Each layer has a finite thickness with defined in- and out-of-plane elastic moduli. The first layer is bounded on the top by air and the last layer is bounded on the bottom by a liquid to mimic corneal *in vivo* conditions. Because collagen fibers are oriented randomly in the equatorial plane, it possesses symmetry across fibers, *i.e.* in the direction normal to the corneal surface, which is mathematically described as transverse isotropy.

#### *1.1 Dimensionless equations of motion in a multilayered NITI material*

Because the macroscopic corneal symmetry did not change following crosslinking, we assume that it contains at least two layers of finite thickness with distinct in- and out-of-plane shear moduli $\mu$ and $G$ respectively. The density of every layer, $\rho$, is assumed to be identical for all layers and equal to that of the liquid bounding the lower layer, $\rho_l = \rho = 1000$ kg·m$^{-3}$. Using Voigt's notation, the stress-strain relationship in each layer can be written as:

$$\begin{bmatrix} \sigma_{xx} \\ \sigma_{yy} \\ \sigma_{zz} \\ \tau_{yz} \\ \tau_{xz} \\ \tau_{xy} \end{bmatrix} = \begin{bmatrix} \lambda + 2\mu & \lambda & \lambda & & & \\ \lambda & \lambda + 2\mu & \lambda & & & \\ \lambda & \lambda & \lambda + \delta & & & \\ & & & G & & \\ & & & & G & \\ & & & & & \mu \end{bmatrix} \begin{bmatrix} \epsilon_{xx} \\ \epsilon_{yy} \\ \epsilon_{zz} \\ \gamma_{yz} \\ \gamma_{xz} \\ \gamma_{xy} \end{bmatrix}, \qquad \text{(S1)}$$

where $\lambda = \rho c_p^2 - 2\mu$ is the Lamé coefficient, and $c_p$ is the speed of the longitudinal wave that ensures incompressibility of the material (the Poisson's ratio of each layer taken individually is $\nu \sim 0.5$).

Newton's second law yields the wave equation of motion in terms of the displacement vector $\vec{u} = (u, v, w)$,

$$\frac{\partial \sigma_{ij}}{\partial x_j} = \rho \frac{\partial u_i}{\partial t^2}. \qquad \text{(S2)}$$

Because our OCE experiments use AμT acting as a pseudo-line source, we can assume a plane-strain state (no displacement polarized along the $y$-axis: $\text{v} = 0$, and no propagation along the $y$-axis: $u = u(x, z), w = w(x, z)$). As such, in the NITI model, the equations of motion can be expressed as:

$$\rho u_{tt} = (\lambda + 2\mu)u_{xx} + G u_{zz} + (\lambda + G)w_{xz}\,, \tag{S3}$$
$$\rho w_{tt} = G w_{xx} + (\lambda + 2\mu)w_{zz} + (\lambda + G)u_{xz}\,, \tag{S4}$$

where the lower indexes indicate derivatives with time $(t)$ or spatial coordinates (x, z).

By introducing the scales:

Position: $x \sim h = \sum_i h_i$ ,
Displacement: $u \sim h$ ,
Time: $t \sim h \cdot \sqrt{\frac{\mu_M}{\rho}}$ ,
Frequency: $f \sim \sqrt{\frac{\mu_M}{\rho}} \cdot \frac{1}{h}$ ,
Wavenumber: $k \sim \frac{1}{h}$,

where $h$ relates to the overall thickness, $h_i$ to the thickness of the$i_{th}$ layer , and $\mu_M = \max(\mu_i)$ is the maximum in-plane shear modulus among the layers. We can then define the dimensionless parameters $t^* = \frac{t}{h}\sqrt{\frac{\mu_M}{\rho}}$ , $u^* = \frac{u}{h}$ , $x^* = \frac{x}{h}$ , $f^* = fh\sqrt{\frac{\rho}{\mu_M}}$ , $k^* = kh$.

By applying this change of variables, we have:

$$\frac{\partial u}{\partial x} = \frac{\partial h \times u^*}{\partial x^*} \times \frac{\partial x^*}{\partial x} = h \cdot \frac{\partial u^*}{\partial x^*} \cdot \frac{1}{h} = \frac{\partial u^*}{\partial x^*}\,, \tag{S5}$$
$$\frac{\partial^2 u}{\partial x^2} = \frac{\partial}{\partial x}\left(\frac{\partial u^*}{\partial x^*}\right) = \frac{1}{h} \cdot \frac{\partial^2 u^*}{\partial {x^*}^2}\,, \tag{S6}$$
$$\frac{\partial^2 u}{\partial t^2} = \frac{\mu_M}{\rho h} \cdot \frac{\partial^2 u^*}{\partial {t^*}^2}\,, \tag{S7}$$

which eventually leads to the dimensionless equations of motion:

$$\rho . u_{tt}^* \cdot \frac{\mu_M}{\rho h} = (\lambda + 2\mu)u_{xx}^* \cdot \frac{1}{h} + G u_{zz}^* \cdot \frac{1}{h} + (\lambda + G)w_{xz}^* \cdot \frac{1}{h}, \tag{S8}$$
$$\rho \mu_M \cdot w_{tt}^* \frac{\mu_M}{\rho h} = G w_{xx}^* \cdot \frac{1}{h} + (\lambda + 2\mu)w_{zz}^* \cdot \frac{1}{h} + (\lambda + G)u_{xz}^* \cdot \frac{1}{h}, \tag{S9}$$

and after rearranging:

$$u_{tt}^* = \beta^2 u_{xx}^* + \alpha^2 u_{zz}^* + \gamma^2 w_{xz}^*\,, \tag{S10}$$
$$w_{tt}^* = \alpha^2 w_{xx}^* + \beta^2 w_{zz}^* + \gamma^2 u_{xz}^*\,, \tag{S11}$$

with

$$\alpha^2 = \frac{G}{\mu_M}, \tag{S12}$$
$$\beta^2 = \frac{(\lambda + 2\mu)}{\mu_M}, \tag{S13}$$
$$\gamma^2 = \frac{(\lambda + G)}{\mu_M}. \tag{S14}$$

For the sake of simplicity, we will later omit the asterisk '$*$' symbol to refer to dimensionless variables.

### *1.2 Dispersion relationship for guided mechanical waves in a multilayered NITI material*

Consider here a material containing multiple layers with identical density, each of finite thickness $h_i$. As such, there is a system of $[2 \times N]$ equations for the $N$ layers:

$$u_{i,tt} = \beta_i^2 u_{xx} + \alpha_i^2 u_{zz} + \gamma_i^2 w_{xz}\ , \qquad \text{(S15)}$$
$$w_{i,tt} = \alpha_i^2 w_{xx} + \beta_i^2 w_{zz} + \gamma_i^2 u_{xz}\ , \qquad \text{(S16)}$$
$$\alpha_i^2 = \frac{G_i}{\mu_N}, \qquad \text{(S17)}$$
$$\beta_i^2 = \frac{(\lambda_i + 2\mu_i)}{\mu_N}, \qquad \text{(S18)}$$
$$\gamma_i^2 = \frac{(\lambda_i + G_i)}{\mu_N}. \qquad \text{(S19)}$$

Assume harmonic solutions of the form:

$$u_i(x,z,t) = A_i e^{i(kx + l_i z - \omega t)}\ , \qquad \text{(S20)}$$
$$w_i(x,z,t) = B_i e^{i(kx + l_i z - \omega t)}\ . \qquad \text{(S21)}$$

Without loss of generality, assume $B_i = 1$. By substituting equations (S20) and (S21) into the equations of motion (S15) and (S16), the constants $l_i$ and $A_i$ can be determined for each frequency and wavenumber such that:

$$l_i = \pm\sqrt{\frac{1}{2}\left[\phi_i \pm \sqrt{\phi_i - 4q_{\alpha_i}^2 q_{\beta_i}^2}\,\right]}\ , \qquad \text{(S22)}$$

$$A_i = \pm\left[-\frac{\sqrt{2}\frac{\gamma_i^2 k}{\alpha_i^2}\sqrt{\phi_i \pm \sqrt{\phi_i - 4q_{\alpha_i}^2 q_{\beta_i}^2}}}{\phi_i + \frac{2\beta_i^2}{\alpha_i^2} q_{\beta_i}^2 \pm \sqrt{\phi_i - 4q_{\alpha_i}^2 q_{\beta_i}^2}}\right], \qquad \text{(S23)}$$

where

$$\phi_i = \frac{\gamma_i^4 k^2}{\alpha_i^2 \beta_i^2} - \frac{\alpha_i^2}{\beta_i^2} q_{\alpha_i}^2 - \frac{\beta_i^2}{\alpha_i^2} q_{\beta_i}^2\ , \qquad \text{(S24)}$$
$$q_{\alpha_i}^2 = k^2 - \frac{\omega^2}{\alpha_i^2}, \qquad \text{(S25)}$$
$$q_{\beta_i}^2 = k^2 - \frac{\omega^2}{\beta_i^2}. \qquad \text{(S26)}$$

The full solutions for a given layer can be expressed as the combination of 4 partial waves:

$$u_i(x,z,t) = \sum_{j=1}^{4} C_{i,j} A_{i,j} e^{il_{i,j}z} e^{i(kx - \omega t)}\ , \qquad \text{(S27)}$$
$$w_i(x,z,t) = \sum_{j=1}^{4} C_{i,j} e^{il_{i,j}z} e^{i(kx - \omega\ )}\ . \qquad \text{(S28)}$$

In the fluid, the dimensionless velocity potential is:

$$\Phi = C_{N,5} e^{\epsilon z} e^{i(kx - \omega t)}\ , \qquad \text{(S29)}$$

where $\epsilon = \sqrt{k^2 - \frac{\omega^2}{\delta^2}}$ and $\delta^2 = \frac{\rho}{\mu_N} c_p^2$ .

The constants $C_{i,j}$ are chosen so that the solutions satisfy the boundary conditions. Traction free air-solid interface sets

$$\sigma_{1,xz} = 0 \quad \text{at} \quad z_1 = 1\ , \qquad \text{(S30)}$$
$$\sigma_{1,zz} = 0 \quad \text{at} \quad z_1 = 1\ . \qquad \text{(S31)}$$

Continuity of normal components of stress and displacement between each layer give

$$\sigma_{i,xz} = \sigma_{i+1,xz} \qquad \text{at} \qquad z_i = 1 - \frac{\sum_{k=1}^{i} h_k}{h}\,, \tag{S32}$$

$$\sigma_{i,zz} = \sigma_{i+1,zz} \qquad \text{at} \qquad z_i = 1 - \frac{\sum_{k=1}^{i} h_k}{h}\,, \tag{S33}$$

$$u_i = u_{i+1} \qquad \text{at} \qquad z_i = 1 - \frac{\sum_{k=1}^{i} h_k}{h}\,, \tag{S34}$$

$$w_i = v_{i+1} \qquad \text{at} \qquad z_i = 1 - \frac{\sum_{k=1}^{i} h_k}{h}\,. \tag{S35}$$

Medium-fluid boundary conditions (zero tangential stress, continuity of normal stress components and speed) set

$$\sigma_{N,xz} = 0 \qquad \text{at} \qquad z_N = 0\,, \tag{S36}$$

$$\sigma_{N,zz} = \sigma_{zz}^{f} \qquad \text{at} \qquad z_N = 0\,, \tag{S37}$$

$$\dot{w}_N = \dot{w}^{f} \qquad \text{at} \qquad z_N = 0\,. \tag{S38}$$

Substituting the general solution into the boundary conditions yields a [4N+1 × 4N+1] homogeneous system for the coefficient: $\mathbf{Mc} = \mathbf{0}$. This system has a nontrivial solution if and only if the determinant of **M** (see Eq. (S39) below) is zero. For a given angular frequency $\omega$, the wavenumber $k$ associated with different wave types (pure shear or guided modes) can be found by minimizing the absolute value of the determinant:

$$M = \begin{bmatrix}
(l_{1,1}A_{1,1}+k)e^{il_{1,1}} & (l_{1,2}A_{1,2}+k)e^{il_{1,2}} & (l_{1,3}A_{1,3}+k)e^{il_{1,3}} & (l_{1,4}A_{1,4}+k)e^{il_{1,4}} \\
[k(\gamma_1^2-\alpha_1^2)A_{1,1}+\beta_1^2 l_{1,1}]e^{il_{1,1}} & k(\gamma_1^2-\alpha_1^2)A_{1,2}+\beta_1^2 l_{1,2}]e^{il_{1,2}} & k(\gamma_1^2-\alpha_1^2)A_{1,3}+\beta_1^2 l_{1,3}]e^{il_{1,3}} & k(\gamma_1^2-\alpha_1^2)A_{1,4}+\beta_1^2 l_{1,4}]e^{il_{1,4}} \\
(l_{1,1}A_{1,1}+k)e^{il_{1,1}z_2} & (l_{1,2}A_{1,2}+k)e^{il_{1,2}z_2} & (l_{1,3}A_{1,3}+k)e^{il_{1,3}z_2} & (l_{1,4}A_{1,4}+k)e^{il_{1,4}z_2} \\
[k(\gamma_1^2-\alpha_1^2)A_{1,1}+\beta_1^2 l_{1,1}]e^{il_{1,1}z_2} & k(\gamma_1^2-\alpha_1^2)A_{1,2}+\beta_1^2 l_{1,2}]e^{il_{1,2}z_2} & k(\gamma_1^2-\alpha_1^2)A_{1,3}+\beta_1^2 l_{1,3}]e^{il_{1,3}z_2} & k(\gamma_1^2-\alpha_1^2)A_{1,4}+\beta_1^2 l_{1,4}]e^{il_{1,4}z_2} \\
A_{1,1}e^{il_{1,1}z_2} & A_{1,2}e^{il_{1,2}z_2} & A_{1,3}e^{il_{1,3}z_2} & A_{1,4}e^{il_{1,4}z_2} \\
e^{il_{1,1}z_2} & e^{il_{1,2}z_2} & e^{il_{1,3}z_2} & e^{il_{1,4}z_2} \\
0 & 0 & 0 & 0 \\
\vdots & \vdots & \vdots & \vdots
\end{matrix}$$

$$\begin{matrix}
0 & 0 & 0 & 0 & \cdots \\
0 & 0 & 0 & 0 & \cdots \\
-(l_{2,1}A_{2,1}+k)e^{il_{2,1}z_2} & -(l_{2,2}A_{2,2}+k)e^{il_{2,2}z_2} & -(l_{2,3}A_{2,3}+k)e^{il_{2,3}z_2} & -(l_{2,4}A_{2,4}+k)e^{il_{2,4}z_2} & \cdots \\
-[k(\gamma_2^2-\alpha_2^2)A_{2,1}+\beta_2^2 l_{2,1}]e^{il_{2,1}z_2} & -[k(\gamma_2^2-\alpha_2^2)A_{2,2}+\beta_2^2 l_{2,2}]e^{il_{2,2}z_2} & -[k(\gamma_2^2-\alpha_2^2)A_{2,3}+\beta_2^2 l_{2,3}]e^{il_{2,3}z_2} & -[k(\gamma_2^2-\alpha_2^2)A_{2,4}+\beta_2^2 l_{2,4}]e^{il_{2,4}z_2} & \cdots \\
-A_{2,1}e^{il_{2,1}z_2} & -A_{2,2}e^{il_{2,2}z_2} & -A_{2,3}e^{il_{2,3}z_2} & -A_{2,4}e^{il_{2,4}z_2} & \cdots \\
-e^{il_{2,1}z_2} & -e^{il_{2,2}z_2} & -e^{il_{2,3}z_2} & -e^{il_{2,4}z_2} & \cdots \\
0 & 0 & 0 & 0 & \cdots \\
\vdots & \vdots & \vdots & \vdots & \vdots
\end{matrix}$$

$$\begin{matrix}
0 & 0 & 0 & 0 & 0 \\
\vdots & \vdots & \vdots & \vdots & \vdots \\
l_{N,1}A_{N,1}+k & l_{N,2}A_{N,2}+k & l_{N,3}A_{N,3}+k & l_{N,4}A_{N,4}+k & 0 \\
k(\gamma_N^2-\alpha_N^2)A_{N,1}+\beta_N^2 l_{N,1} & k(\gamma_N^2-\alpha_N^2)A_{N,2}+\beta_N^2 l_{N,2} & k(\gamma_N^2-\alpha_N^2)A_{N,3}+\beta_N^2 l_{N,3} & k(\gamma_N^2-\alpha_N^2)A_{N,4}+\beta_N^2 l_{N,4} & \frac{\omega\rho^f}{\rho} \\
\omega & \omega & \omega & \omega & -i\epsilon
\end{bmatrix}. \tag{S39}$$

## Supplementary Material 2:

### Effective $A_0$ guided mode for a multilayered NITI medium

#### *2.1 Do engineering moduli correctly define the effective guided mechanical wave in a multi-layered NITI medium?*

As discussed in Section 3.3 of the main manuscript, elastic moduli determined for all layers of CXL-treated cornea can be used to compute effective corneal engineering moduli, where $\mu_{eff}$ uses a simple mixture rule (Eq. (5) of the main manuscript) whereas $G_{eff}$ requires the inverse mixture rule (Eq. (4) of the main manuscript).

We also checked whether this model can describe guided wave behavior in the partially crosslinked cornea when considered as an effective homogeneous material. First, we applied the fitting procedure to the treated cornea, considering it as a single layer with 'effective' moduli (see Fig. S1). We could determine a pair of moduli for the computed *k-f* spectrum that best fit the $A_0$-mode. These effective 'guided wave' elastic moduli were $G_{guided} = 127.5 \mp (12, 17)$ kPa and $\mu_{guided} = 9.3 \mp (8, 18)$ MPa, with a 0.953 goodness of fit.

These values clearly do not equal the corneal effective engineering moduli, $G_{eff} = 77.3 \mp (6{,}10)$ kPa and $\mu_{eff} = 15.4 \mp (8{,}11)$ MPa, computed with the mixture rules described by Eqs. (4), (5) using moduli measured with OCE in both layers. That is, it appears that the effective guided wave behavior in a multi-layered NITI medium is not described by the corneal effective engineering moduli which determine its low-frequency quasi-static deformation.

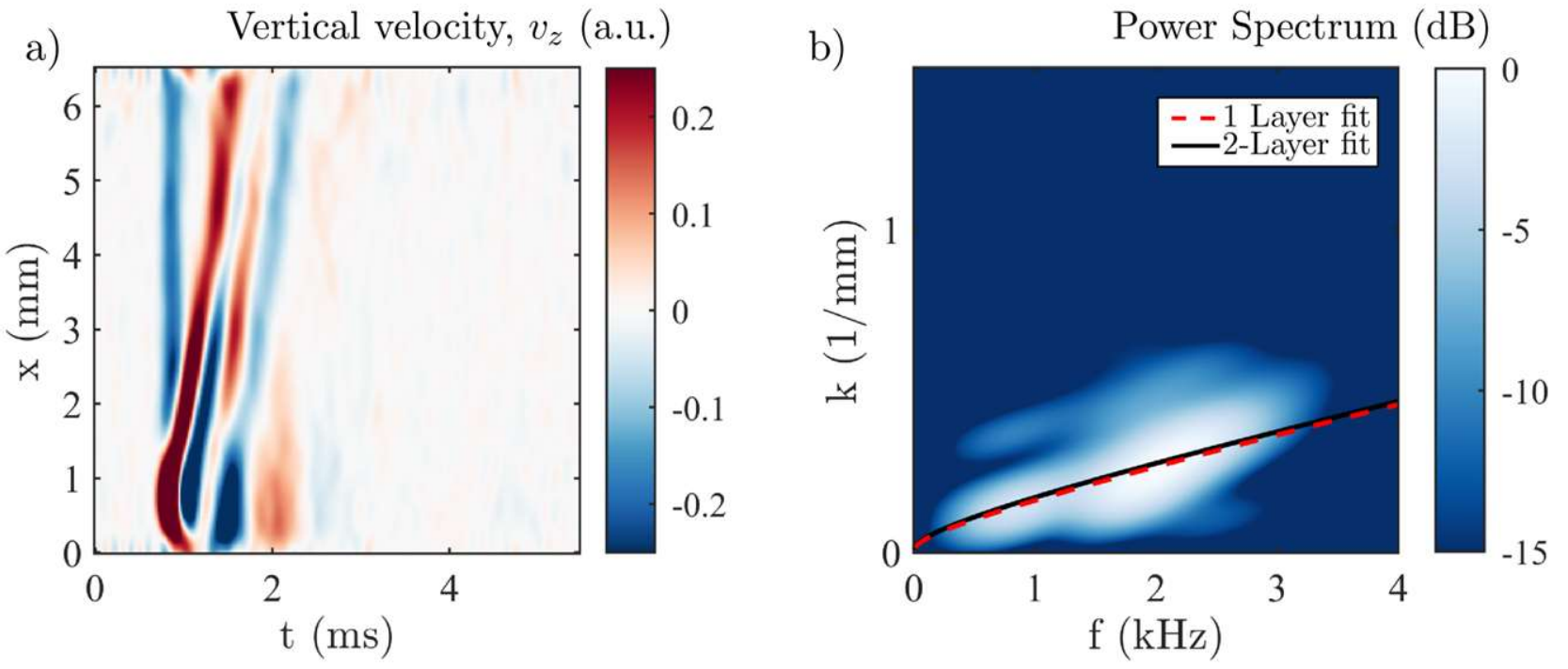

Figure S1. Fitting experimental results obtained for a partially (in-depth) CXL-treated cornea with either 1- or 2-layer analytical model. a) Measured vertically polarized top-surface vibration velocity field (*x-t* plot) of the guided wave. Its 2D-FFT spectrum with best-fit dispersion curves for 1-layer (red dashed line) and 2-layer (black solid line) models superimposed.

#### *2.2 A single-layer approximation for the computation the guided mechanical wave dispersion in a multi-layered NITI medium*

Using our analytical model, we further investigated the accuracy of the mixture rule (Eqs. (4) and (5)) to predict the effective 'guided wave' behavior. We considered different distributions of stiffness: i) a two-layer case, as assumed for CXL-treated corneas; ii) a five-layer case with random distribution of stiffness and thickness of the layers; iii) a medium with both G and μ following an exponential decay in stiffness from top to bottom. In all cases, the total medium thickness was h=500 μm and its top and bottom layers replicated the corneal boundary

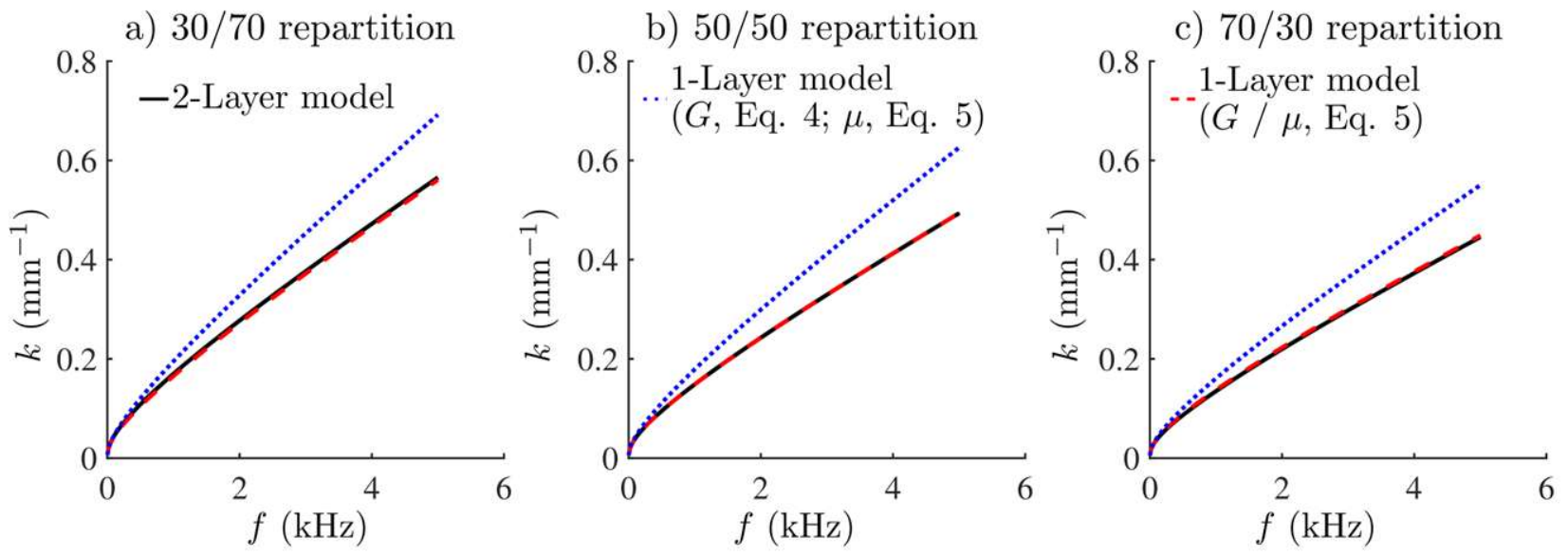

Figure S2. Computation of $A_0$-mode dispersion spectrum using different models and sets of mechanical moduli. In all 3 panels, black solid line corresponds to the exact solution obtained with the 2-layer model; blue dotted line corresponds to a 1-layer model with effective engineering mechanical moduli (Eq. 4 for $G$ and Eq. 5 for $\mu$); red dashed line corresponds to a 1-layer model with both effective moduli computed with Eq. 5.
Three different thicknesses of the top layer are considered a) 150 $\mu$m (30/70 repartition), b) 250 $\mu$m (50/50 repartition), and c) 350 $\mu$m (70/30 repartition). Moduli of the top layer are $G_{ant}$ = 300 kPa and $\mu_{ant}$ = 30 MPa and for the bottom layer are $G_{pos}$ = 60 kPa and $\mu_{pos}$ = 5 MPa, and the total thickness is $h$ = 500 μm for all 3 cases.

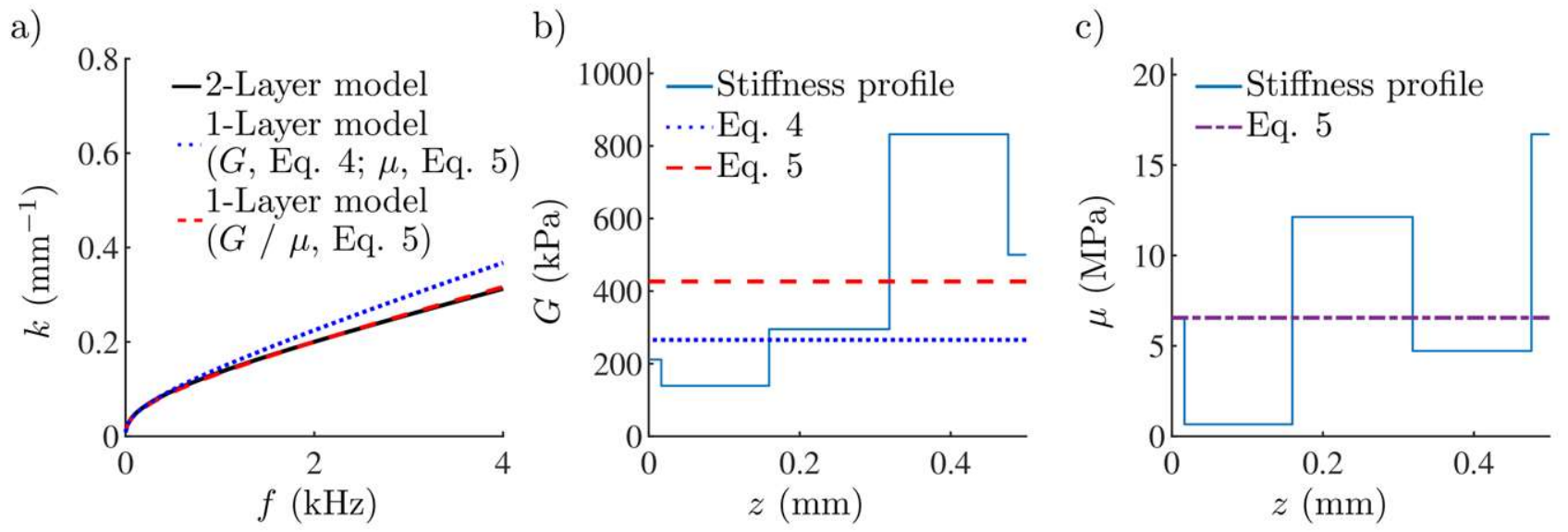

Figure S3. Computation of $A_0$-mode dispersion spectrum in a 5-layer medium using different models and sets of mechanical moduli. In panel a), black solid line corresponds to the exact solution obtained with the 2-layer model; blue dotted line corresponds to a 1-layer model with effective engineering mechanical moduli (Eq. 4 for $G$ and Eq. 5 for $\mu$); red dashed line corresponds to a 1-layer model with both effective moduli computed with Eq. 5. b) Distribution of modulus $G$ in the 5-layer medium (light blue solid line), and its effective value obtained with Eq.4 (blue dotted line) and with Eq.5 (red dashed line). c) Distribution of modulus $\mu$ in the 5-layer medium (light blue solid line), and its effective value obtained with Eq.5 (purple dash-dotted line). Total thickness of the medium is $h$ = 500 μm.

conditions. The Matlab scripts to reproduce these results are provided in Supplementary Material.

Results for case i) are presented in Fig. S2 for a material containing two layers with $G_{ant}$ = 300 kPa and $\mu_{ant}$ = 30 MPa, and $G_{pos}$ = 60 kPa and $\mu_{pos}$ = 5 MPa. For this case, we also explore different ratios between thicknesses of anterior and posterior layers. Results show that the $A_0$-mode computed with the effective engineering moduli using the mixture rules of Eqs. (4) and (5) does not match the exact analytical solution computed using the individual stiffness moduli of the layers, *i.e.*, using the 2-layered model directly. The difference is especially pronounced in the high-frequency range.

Results for case ii) are presented in Fig. S3. The distribution of stiffness for respectively $G$ and $\mu$ are shown in Figs. S3 b,c). The thicknesses of layers were random with the total thickness of $h$ = 500 $\mu$m. As for case i), we also see that assuming a single layer material with averaged moduli computed using engineering effective mechanical moduli does not accurately predict $A_0$-mode dispersion.

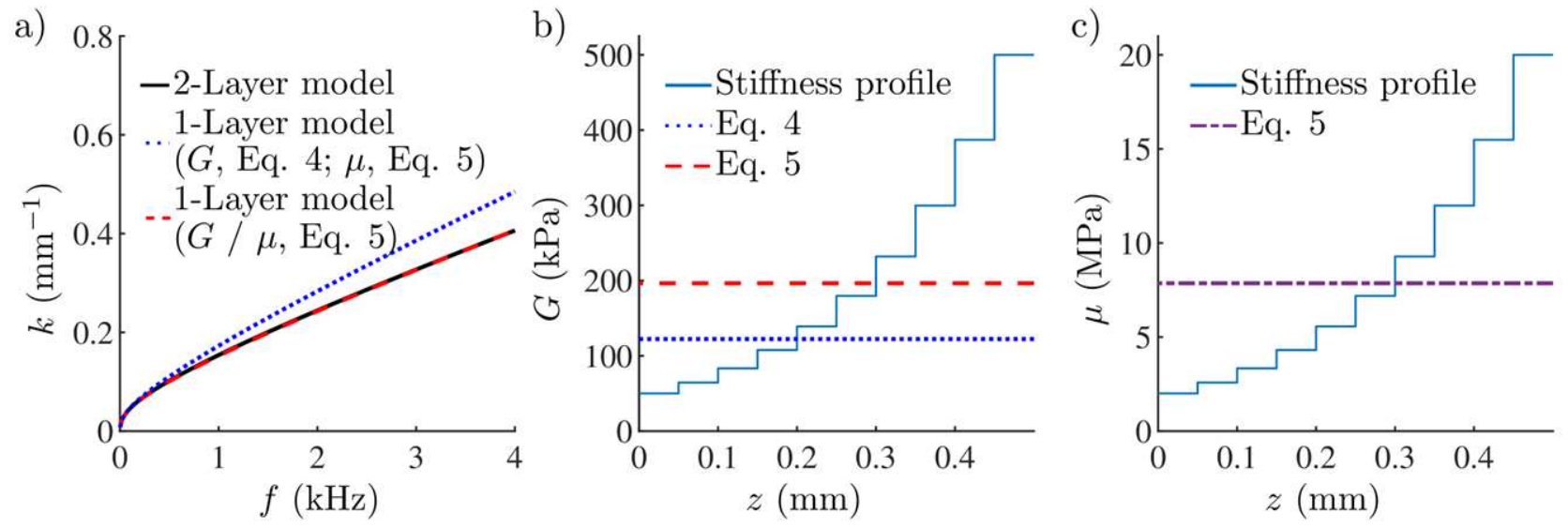

Figure S4. Computation of $A_0$-mode dispersion spectrum in a 10-layer medium with an exponential decay of elastic moduli using different models and sets of mechanical moduli. In panel a), black solid line corresponds to the exact solution obtained with the 2-layer model; blue dotted line corresponds to a 1-layer model with effective engineering mechanical moduli (Eq. 4 for $G$ and Eq. 5 for $\mu$); red dashed line corresponds to a 1-layer model with both effective moduli computed with Eq. 5. b) Distribution of modulus $G$ in the 5-layer medium (light blue solid line), and its effective value obtained with Eq.4 (blue dotted line) and with Eq.5 (purple dash-dotted line). c) Distribution of modulus $\mu$ in the 5-layer medium (light blue solid line), and its effective value obtained with Eq.5 (red dashed line). Total thickness of the medium is $h = 500$ μm.

Results for case iii) are presented in Fig. S4. The distribution of stiffness for respectively $G$ and $\mu$ are shown in Figs. S4 b, c). We considered 10 layers where $G$ decreased exponentially from 500 kPa on top to 50 kPa at the bottom and $\mu$ decreased exponentially as well from 20 MPa on top to 2 MPa at the bottom. As for both cases i) and ii), the effective mechanical moduli do not describe the dispersion curve accurately, especially over the high-frequency range.

Thus, we can conclude that the mixture rules providing effective engineering moduli of multi-layered NITI media do not accurately describe guided wave behavior. As such, reconstruction of effective moduli from OCE measurements in a partially CXL-treated cornea should be performed with care.

The second question is which mixture model would describe guided wave behavior in a partially CXL-treated cornea with reasonable accuracy. This is an open and non-trivial question outside the scope of this paper. However, as shown in Figs. S2, S3 and S4, a simple direct mixture rule for both in- and out-of-plane moduli (Eq. (5)) produces a dispersion curve closely matching the N-layer model.

For the two-layer case (Fig. S2), equal thickness layers (50/50 split ratio, Fig. S2b) show a near-exact match, while the larger the difference between the layer thicknesses, the larger the difference between the solutions, although this difference remains small. In Fig. S3 (random thickness distribution), the simple mixing rule provides a reasonably good match, even though there is a small difference in the dispersion curves. In Fig. S4 (exponential decay of moduli), there is an almost exact match between the single layer effective model and the N-layer case.

Based on this observation, we assume that a simple mixture rule applied to both mechanical moduli $\mu$ and $G$ provides a reasonable description of guided wave dispersion in multi-layer NITI materials. The effective guided model and the exact N-layer solution only differ slightly for an unequal distribution of thicknesses. However, we note that this rule is empirical and is not absolutely accurate. Finding an exact analytical solution may be difficult and include frequency-dependent terms. This is a subject for future studies.